\documentclass{appolb}
\usepackage{epsfig}

\begin{document}
\title{SIGNATURES OF FIREBALL FRAGMENTATION\\ AT THE PHASE TRANSITION%
\thanks{Presented at Excited QCD 2010, Jan 31 -- Feb 6, 2010, Tatransk\'a Lomnica, Slovakia}%
}
\author{Boris Tom\'a\v{s}ik
\address{FNSPE, Czech Technical University in Prague, Prague, Czech Republic\\
and Univerzita Mateja Bela, Bansk\'a Bystrica, Slovakia }
}
\maketitle

\begin{abstract}
It is explained why and how the fireball created in ultrarelativistic 
nuclear collisions can fragment when passing the phase transition. 
It can happen at the first-order phase transition but is not excluded 
even at high collision energies where the smooth crossover is present. 
Two potential observables sensitive to the appearance of fragmentation 
are reviewed: event-by-event changes of rapidity distributions and 
proton correlation in relative rapidity.
\end{abstract}
\PACS{25.75.-q,25.75.Dw,25.75.Gz}


\section{Introduction}

In ultrarelativistic nuclear collisions we probe bulk strongly 
interacting matter at extreme temperatures. The medium is hottest 
and densest immediately after the passage of one nucleus through the 
other.  In collisions at highest 
available energies the deconfined state of colour charges is created
shortly after the initial impact. Then, the subsequent expansion and cooling 
is rather 
fast. In the longitudinal direction it is described by a Hubble-like scaling 
flow profile. The flow velocity in transverse direction reaches 
about 0.7$c$ at $\sqrt{s}_{NN} = 200$~GeV. The total lifetime 
of the hot fireball is about 10~fm/$c$. 

If the fast evolution of the fireball includes the transition from 
deconfined to hadronic state we have a reason 
to expect non-equilibrium phenomena. I want to argue that 
one such scenario which could be realized is that the fireball matter 
violently decays into many smaller fragments or droplets \cite{Torrieri:2007fb}. In other works 
this process may be called cavitation or fragmentation \cite{Rajagopal:2009yw}.

In the next Section arguments are presented why fragmentation at the transition 
is a thinkable process. If it really happens, however, we ought to identify 
observables that are sensitive to it. In order to study such observables,
a Monte Carlo model called DRAGON has been developed for the creation 
of artificial data including the effect of fragmentation. It is briefly 
introduced in Section \ref{s:dragon}. In Section \ref{s:obser} I explain 
two potential observables sensitive to the fragmentation scenario: event-by-event
fluctuations of rapidity distributions and correlation functions in relative 
rapidity.


\section{Fragmentation of the fireball: why and when}

When bulk matter passes through a phase transition fast, this can 
lead to fragmentation \cite{Csorgo:1994dd,Mishustin:1998eq}. 
Let us first explain this on an example 
of a first order phase transition. For illustration, let us consider 
an isotherm of the van der Waals equation of state. Below the critical 
temperature it exhibits a wiggle which is connected with the phase 
transition. Maxwell rule dictates a horizontal line instead 
of a wiggle. This describes a \emph{slow} evolution of the system.
Along the straight line gradually larger and larger volume switches
to the new phase. When isothermal expansion is \emph{fast}, the system first 
keeps to the original isotherm and returns to the straight horizontal 
line when fluctuations initiate the phase transition. In case of a 
\emph{very fast} expansion even the local minimum of the wiggle 
can be reached. Beyond this point the system becomes mechanically 
unstable and spinodal fragmentation sets in \cite{Mishustin:2001re}. 
Realistic 
estimates show that in nuclear collisions the expansion rate may be 
larger than the nucleation rate for the bubbles of the new phase and 
the spinodal fragmentation scenario could be realistic \cite{Scavenius:2000bb}. 

Nevertheless, at collision energies above few tens of GeV per nucleon
the baryochemical potential is so low that a first order phase transition
appears highly unlikely and we rather see a  smooth crossover.
Spinodal fragmetnation is then irrelevant. It has been noted recently, 
however, that the bulk viscosity has a sharp peak at $T_c$ 
as a function of temperature \cite{prattbulk,kharbulk,Meyer:2007dy}. 
Based on this observation it has been 
proposed that the fireball could fragment at $T_c$ as a consequence 
of the expansion flow which is established at this point already, and
the sudden unwillingness of the system to change the volume, connected 
with the appearance of the bulk viscosity \cite{Torrieri:2007fb}. 

In summary, fragmentation of the fireball at $T_c$ is a realistic scenario
at any ultrarelativistic collision energy although it may be more natural 
to appear at higher baryochemical potential connected with first order 
phase transition.


\section{DRAGON: the Monte Carlo model}
\label{s:dragon}

In order to test various observables that can be sensitive to the production 
of hadrons from droplets a Monte Carlo model was constructed, called DRoplet
an hAdron GeneratOr for Nuclear collisions (DRAGON) \cite{dragon}. The model includes two 
types of particle production: final state hadrons may be emitted directly 
from the bulk fireball or from droplets into which (part of) the fireball 
decayed. These two sources can be combined since even after fragmentation 
into smaller droplets some dilute matter may remain in the space between 
them. The shape and expansion pattern of the fireball from which the droplets
originate are inherited from the blast wave model. Droplets obtain the 
velocity which is given by the local flow velocity at the position where
they are produced. Their sizes can be set in the simulation. Generally, they 
should be given by the expansion gradients and properties of the medium. 
The chemical composition of the produced hadrons is determined according to 
equilibrium prescription. Mesons and baryons with masses up to 1.5 and 
2~GeV/$c^2$ are included, respectively. Resonance decays are accounted 
for.


\section{Observables sensitive to fragmentation}
\label{s:obser}

In general, a decay of the fireball into many droplets will be 
reflected in clustering in momentum distribution. Velocities of the 
produced hadrons will be clustered around the velocity of the droplet 
from which they originate. Thus we expect clustering in the momentum space. 
The amount by which the momenta of hadrons differ from the vector corresponding
to the droplet velocity increases with growing temperature and is smaller 
for heavier hadrons. For pions, however, clusters are well visible in 
the momentum space if the temperature is unrealistically low, e.g.\ 10~MeV. 
For realistic freeze-out temperature the momenta are smeared. Clustering
is more pronounced, however, for heavier particles. 

In the following we mention two potential candidate observables for the 
identification of clustering in momentum space that could be due to droplets.

\subsection{Fluctuations of rapidity distributions}
\label{s:KS}

Suppose we select very narrow centrality class and consider a group of 
events with initial conditions as close to each other as possible. In such 
a case one might expect the same scenario running in each event. If the fireball 
stays in one piece, we would then expect that the spectra of final state hadrons
will be identical within statistical uncertainties. Not so, however, in the case of 
fireball fragmentation. In this case, droplets will be produced in different 
places in each event and consequently the momentum clusters will also 
differ event by event. We further focus on rapidity distributions. The 
statement is that in case of fragmentation there will be non-statistical 
differences between rapidity spectra measured individually in each event.

The statistical tool which can be used for identification of such a situation 
is the Kolmogorov-Smirnov test \cite{kolm,smir,Melo:2009xh}. It defines the measure of ``unlikeness'' of 
two empirical distributions, which in our case will be two samples of measured
hadron rapidities from one event. For a pair of events a quantity $Q$ is defined.
It is the conditional probability that two events will look more differently, 
provided they are generated from the same probability distribution. It is 
constructed so that a sample of events with common underlying probability 
(e.g.\ rapidity) density will produce a flat histogram of $Q$'s, if $Q$ is
measured in a large number of \emph{event pairs}. If within a set of events 
we obtain too many $Q$ values close to 0, then this means that the events are 
not drawn independently from the same probability distribution. 

\begin{figure}
\centerline{\epsfig{file=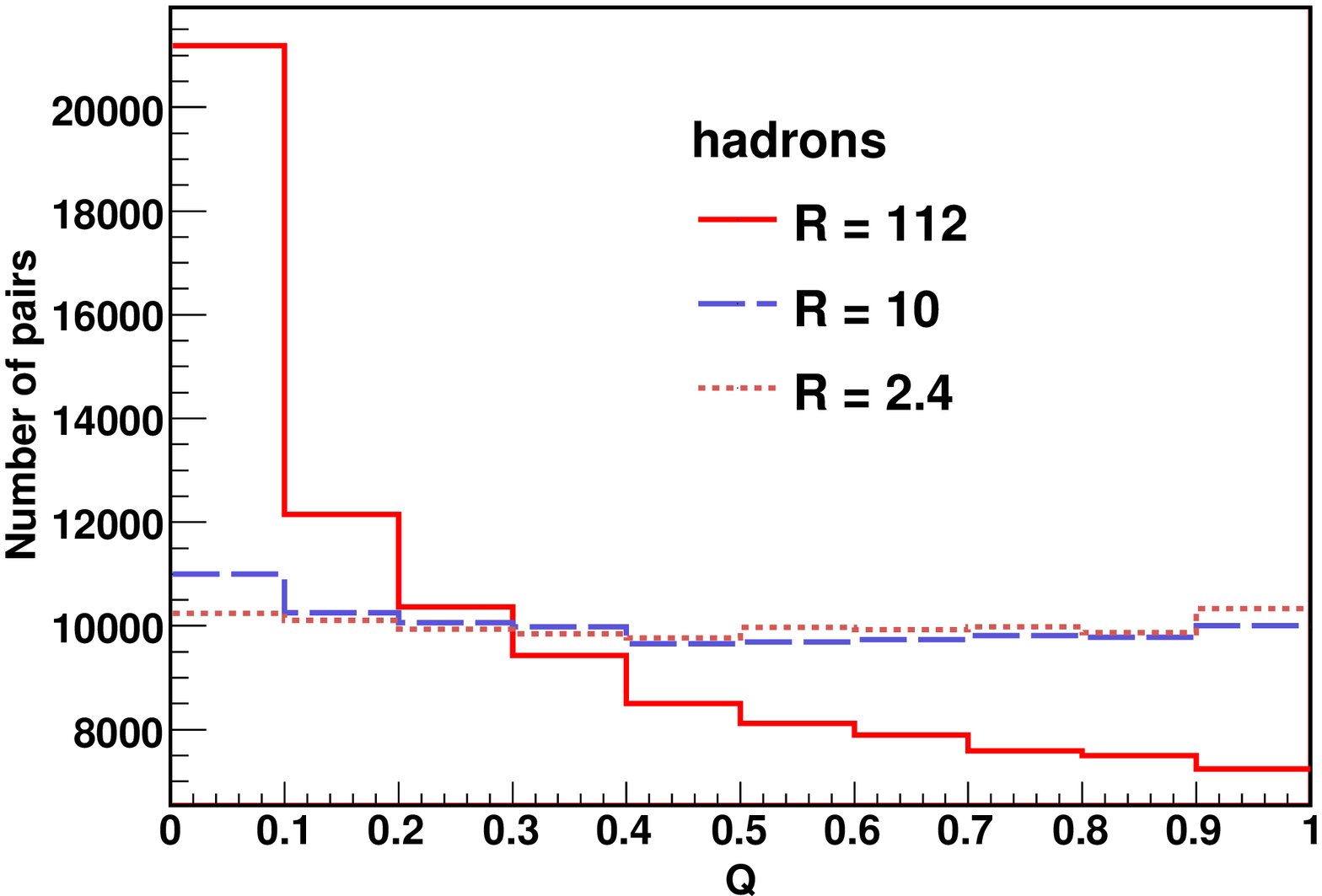,width=0.48\textwidth}
\epsfig{file=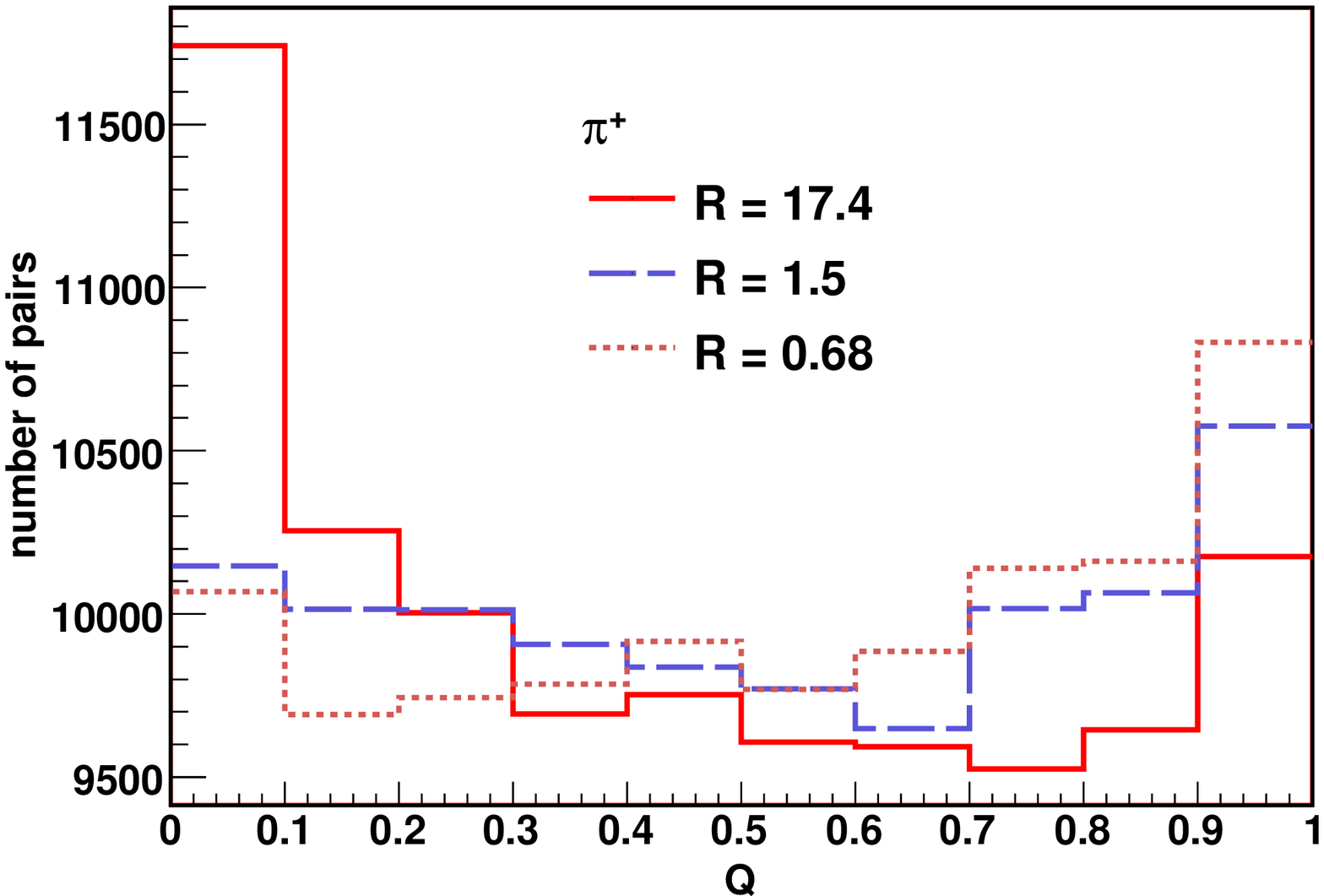,width=0.48\textwidth}}
\caption{The $Q$-histograms evaluated on 10$^5$ event pairs taken from 
samples of 10$^4$ events. The red solid line shows the histogram from events with 
droplet production. The dashed and dotted lines show results from simulations with only 
bulk production assumed from $\sqrt{s_{NN}} = 130$~GeV and 9~GeV, respectively. 
Left panel: all charged hadrons, right panel: only positive pions. $R$ gives the 
statistical significance of the peak at $Q=0$.
\label{f:ksfig}
}
\end{figure}

We have simulated sets of events with DRAGON. In order to judge on the effect 
of fragmentation we investigated events where all hadrons have been emitted from 
droplets. The average size of droplets was 5~fm$^3$. For comparison, other sets of data 
were simulated where no droplets have been taken into account. Chemical composition 
was tuned as to correspond to that at 
$\sqrt{s_{NN}} = 130$~GeV and the rapidity distribution of hadrons or droplets 
was uniform. One simulation without droplets was performed with Gaussian rapidity 
distribution and chemical composition from collisions at $\sqrt{s_{NN}} = 9$~GeV
(FAIR energy). In Figure~\ref{f:ksfig}
we clearly see that the presence of droplets leads to a pronounced peak at low $Q$. 
In charged hadrons there is a small peak also in case with no droplets. This 
is due to resonance decays which act like very small droplets: they correlate
two or three final state hadrons. To get rid of them one can use only pions 
or only protons for the measurement. On the other hand, that decreases the statistics 
and has an impact on the method, which is, strictly speaking, so far 
only well constructed in the limit of very large multiplicity.


\subsection{Proton rapidity correlations}
\label{s:rapcor}

Two hadrons steming from the same droplet will have similar rapidities. 
Thus we expect non-trivial correlation function in case of fragmentation. 
The hadron rapidities will differ from the rapidity of the droplet by thermal 
component of the velocity. Thermal motion is slower for heavier particles. Thus 
it is reasonable to choose heavy particles. Since the abundance of hadrons 
of given kind decreases with increasing mass, we have to find a balance between 
small thermal smearing and large statistics. Protons appear as a good choice \cite{pratt94,randrup05}. 

\begin{figure}
\centerline{\epsfig{file=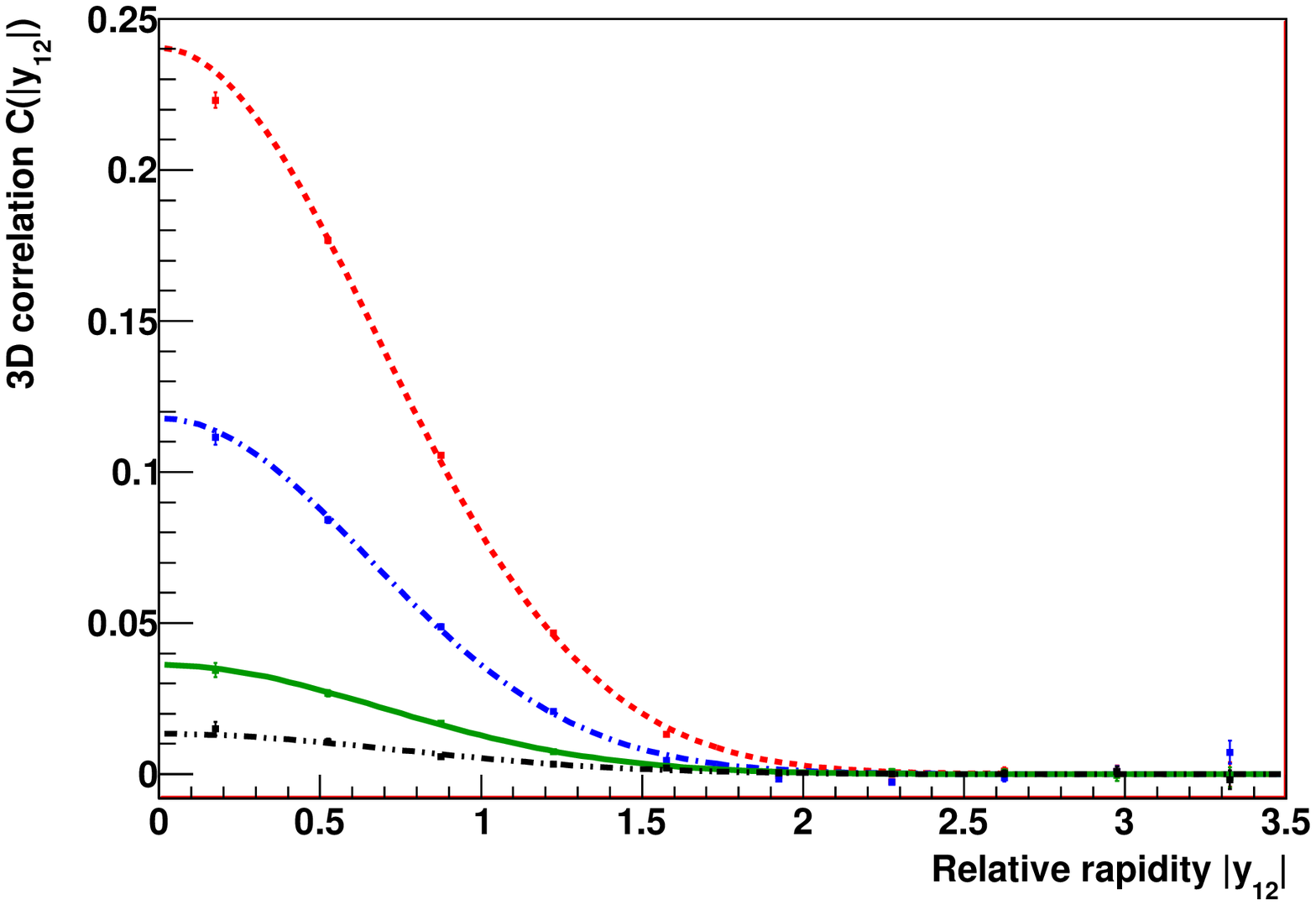,width=0.5\textwidth}
\epsfig{file=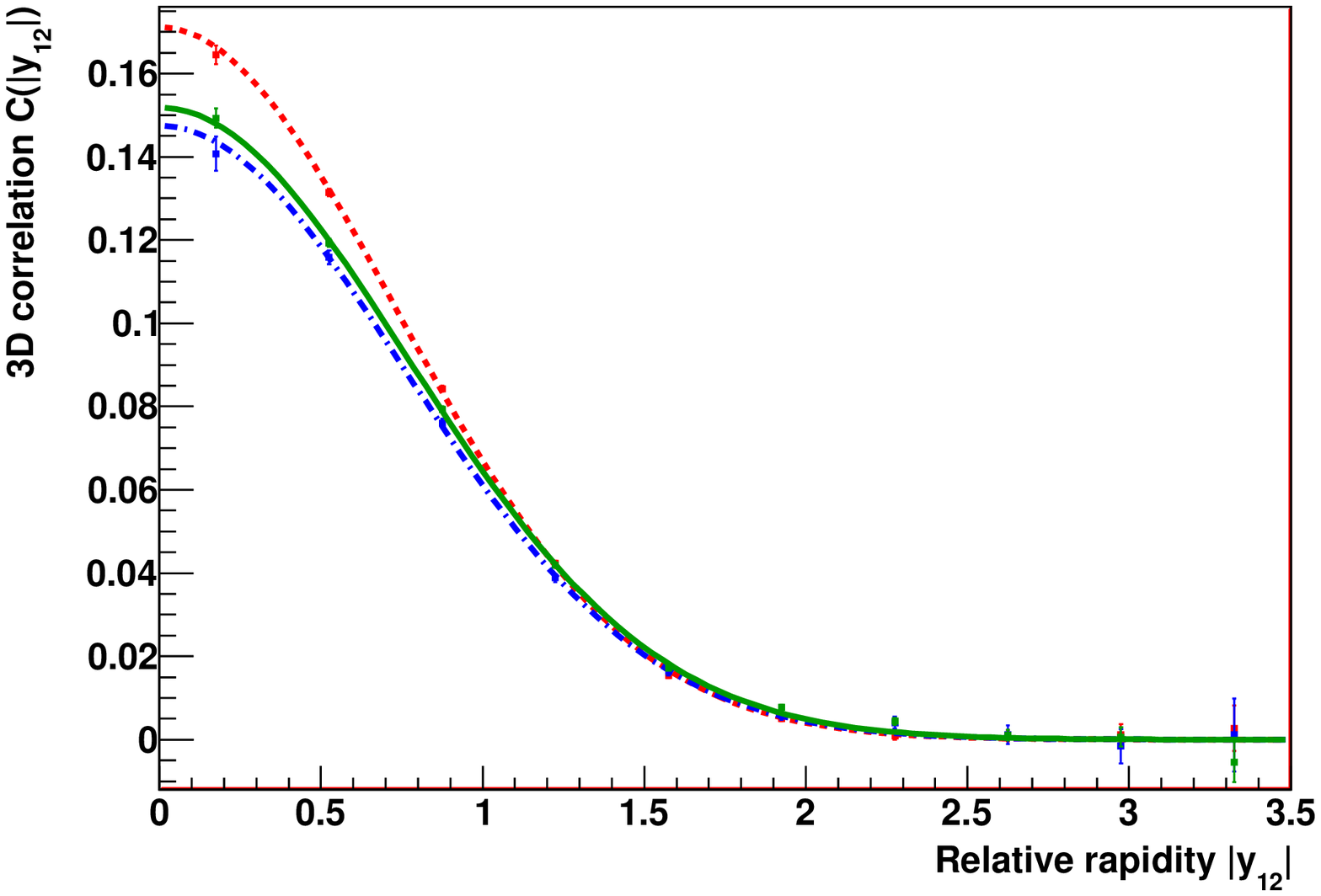,width=0.5\textwidth}}
\caption{
Correlation functions of protons in relative rapidity  calculated for a situation 
in Au+Au collisions at $\sqrt{s_{NN}} = 130$~GeV. Left panel: one half of hadrons 
is produced from droplets, their average sizes are $5\,\mbox{fm}^3$ (black dash-dot-dotted curve), 
$10\,\mbox{fm}^3$ (green solid), $25\,\mbox{fm}^3$
(blue dash-dotted), and $50\,\mbox{fm}^3$ (red dotted). Right panel: the influence of resonance 
decays on the correlation function. All particles form droplets with average size $25\,\mbox{fm}^3$. 
Correlation function with all resonance decays 
included (red dotted), simulation with no resonance production 
included (green solid), simulation with resonances included but 
protons from decays of $\Delta$ resonances not taken into analysis
(blue dash-dotted).
\label{f:pcor}
}
\end{figure}

On DRAGON-generated events we have studied proton correlations in 
\emph{relative rapidity} defined as 
$y_{12} = \ln  [ \gamma_{12} + \sqrt{\gamma_{12}^2 - 1}  ]$ where
$\gamma_{12} = \frac{p_1\cdot p_2}{m_1 m_2}$ \cite{Schulc:2010hj}.
We found that the correlation function is measurably non-zero for droplets
with average volume as small as 5~fm$^3$ and also if only a part of the hadrons
comes from the droplets and the rest is from the bulk in between. Surprisingly, 
we observed that resonances which decay into protons make the proton 
correlation function even more pronounced (Figure~\ref{f:pcor}). 
This is due to their higher mass resulting in weaker thermal smearing.


\section{Conclusions}

We have presented two possible observables which could help to indicate fragmentation 
of the fireball in ultrarelativistic nuclear collisions: event-by-event 
changes of the rapidity distributions identified via Kolmogorov-Smirnov test and
correlation of protons in relative rapidity. Among other observables which 
are worth investigating there are imaging, elliptic flow and its fluctuations, 
correlations in rapidity and azimuthal angle, etc. We plan to address
these topics in the near future.


\subsubsection*{Acknowledgemens}
I thank my collaborators with whom the results presented here were obtained: 
M.\ Bleicher, M.\ Gintner, S.\ Kor\'ony, I.\ Melo, I.\ Mishustin, M.\ Schulc, 
S.\ Vogel, G.\ Torrieri. I thank the organisers for the invitation to this 
conference. Supported in parts 
by grants No.~MSM 6840770039,
LC 07048 (Czech Republic), VEGA 1/4012/07 (Slovakia) and by DAAD.

\end{document}